\documentclass[aps,prl,twocolumn,showpacs,preprintnumbers,amsmath,amssymb,floatfix]{revtex4-2}

\usepackage{graphicx}
\usepackage{longtable}
\usepackage{dcolumn}
\usepackage{bm}
\usepackage{color}
\usepackage[normalem]{ulem}
\usepackage[colorlinks=true, urlcolor=blue, linkcolor=blue, citecolor=blue]{hyperref}
\usepackage[all]{hypcap}
\newcommand{\beq}{\begin{equation}}
\newcommand{\eeq}{\end{equation}}
\newcommand{\bea}{\begin{eqnarray}}
\newcommand{\eea}{\end{eqnarray}}

\begin{document}

\title{\textit{d}-wave Surface Altermagnetism in Centrosymmetric Collinear Antiferromagnets}

\author{Ersoy \c{S}a\c{s}{\i}o\u{g}lu$^{1}$}\email{ersoy.sasioglu@physik.uni-halle.de}
\author{Ingrid Mertig$^{1}$}
\author{Samir Lounis$^{1}$}

\affiliation{$^{1}$Institute of Physics and Halle-Berlin-Regensburg Cluster of Excellence CCE, Martin Luther University Halle-Wittenberg, 06120 Halle (Saale), Germany}

\date{\today}

\begin{abstract}
Broken inversion symmetry at the surfaces of centrosymmetric collinear antiferromagnets lifts
combined inversion and time-reversal symmetry ($PT$) and can, in principle, enable
nonrelativistic \textit{d}-wave spin splitting, termed surface altermagnetism. Combining symmetry
analysis with first-principles calculations, we show that surface inversion breaking, while
necessary, is not sufficient for this effect. Surface altermagnetism emerges only when no
antiunitary symmetry survives at the surface that exchanges the two antiferromagnetically coupled
surface sublattices and enforces spin degeneracy. We demonstrate this mechanism explicitly for
the centrosymmetric \textit{G}-type antiferromagnets V$_3$Al and BaMn$_2$Sb$_2$, and contrast
it with MnPt, where a sublattice-exchanging symmetry survives at the surface in the form of
translation--time-reversal symmetry ($tT$), thereby preserving spin degeneracy despite broken
inversion symmetry. The mechanism is shown to apply across multiple classes of centrosymmetric 
antiferromagnets and remains robust against spin--orbit coupling, although relativistic spin 
mixing in heavier-element compounds may reduce the observable spin polarization. These results 
establish a symmetry-based route toward realizing robust nonrelativistic momentum-dependent 
spin polarization at antiferromagnetic surfaces and interfaces.
\end{abstract}

\maketitle

Antiferromagnets (AFMs) possessing combined inversion and time-reversal symmetry  ($PT$)  exhibit
spin-degenerate electronic bands in the absence of spin--orbit coupling (SOC).  The recent discovery
of altermagnetism has challenged this paradigm by demonstrating that collinear compensated magnets
can host large nonrelativistic spin splittings when their magnetic space groups break $PT$ while
preserving the equivalence of opposite-spin sublattices through crystal rotations or mirror
operations \cite{Zunger,Smejkal2020_SciAdv,Smejkal2022_PRX,vsmejkal2022anomalous}. The resulting 
momentum-dependent spin polarization follows characteristic \textit{d}-, \textit{g}-, or \textit{i}-wave 
symmetries dictated by the underlying lattice and has been realized in several classes of transition-metal
compounds \cite{Guo2023_MatTodPhys,Mazin2023_MnTe,Jungwirth2025_AltermagSpintronics,Giuli2024_NodalReview,Feng2022_NatElectron,Osumi2024_PRB}.
Experimentally, altermagnetic band splittings have now been directly resolved in $\alpha$-MnTe,
CrSb, and the metallic room-temperature \textit{d}-wave altermagnet KV$_2$Se$_2$O using
angle-resolved photoemission spectroscopy and complementary transport probes,
establishing altermagnets as a distinct class of compensated magnets with nonrelativistic spin
polarization \cite{Lee2024_PRL_MnTe,Reimers2024_NatCommun_CrSb,Jiang2025_NatPhys_KV2Se2O,biniskos2025systematic}.
Motivated by this discovery, altermagnetism has rapidly evolved into an active research field, 
encompassing symmetry-based classifications, large-scale materials screening, and first-principles 
studies of bulk and low-dimensional systems \cite{yuan2020giant,sodequist2024two,brekke2023two,xu2026chemical,jungwirth2026symmetry,wan2025high,chang2025inverse,bhattarai2025high,Brahimi2024_RuO2_film,bai2024altermagnetism,Tamang_Review_Altermagnets,matsuda2025multiferroic,mcclarty2024landau,sivianes2025optical,jo2025weak,song2025electrical,fukaya2025superconducting}.

An important and largely unexplored question is whether analogous altermagnetic spin splitting can
also emerge at the \emph{surfaces} of otherwise $PT$-symmetric AFMs. While recent
theoretical and materials-driven studies have firmly established altermagnetism in bulk and
low-dimensional periodic systems, the role of surfaces, where translational symmetry is explicitly
broken, remains poorly understood. Recent theoretical work has begun to examine surface electronic
structures of intrinsically altermagnetic bulk materials \cite{sattigeri2023altermagnetic};
however, the emergence of altermagnetic spin splitting at the surfaces of bulk $PT$-symmetric
AFMs, where altermagnetism is absent in the bulk, has not been established.
Surface termination inherently breaks inversion symmetry and thus lifts $PT$ even in
centrosymmetric \textit{C}- and \textit{G}-type AFMs whose bulk bands are strictly
spin-degenerate. Unlike relativistic Rashba or Dresselhaus effects, which originate from SOC in 
noncentrosymmetric crystals \cite{Rashba1960_SovPhysSolidState,Dresselhaus1955_PhysRev}, the 
surface-induced spin splitting discussed here emerges already in the absence of SOC from the local 
antiferromagnetic exchange field. SOC may subsequently modify the detailed surface electronic structure 
through spin mixing and additional relativistic splittings, but it is not the primary microscopic 
origin of the effect. Unlike conventional ferromagnetic exchange splitting, the 
resulting nonrelativistic spin polarization exhibits a symmetry-enforced momentum-dependent sign 
change in momentum space. Whether surface-induced inversion breaking alone is sufficient to 
generate a symmetry-dictated momentum-dependent spin polarization analogous to that of bulk 
altermagnets, or whether additional  degeneracy-protecting symmetries remain operative at the 
surface, therefore remains an open question.

\begin{figure*}[t]
\centering
\includegraphics[width=0.9999\textwidth]{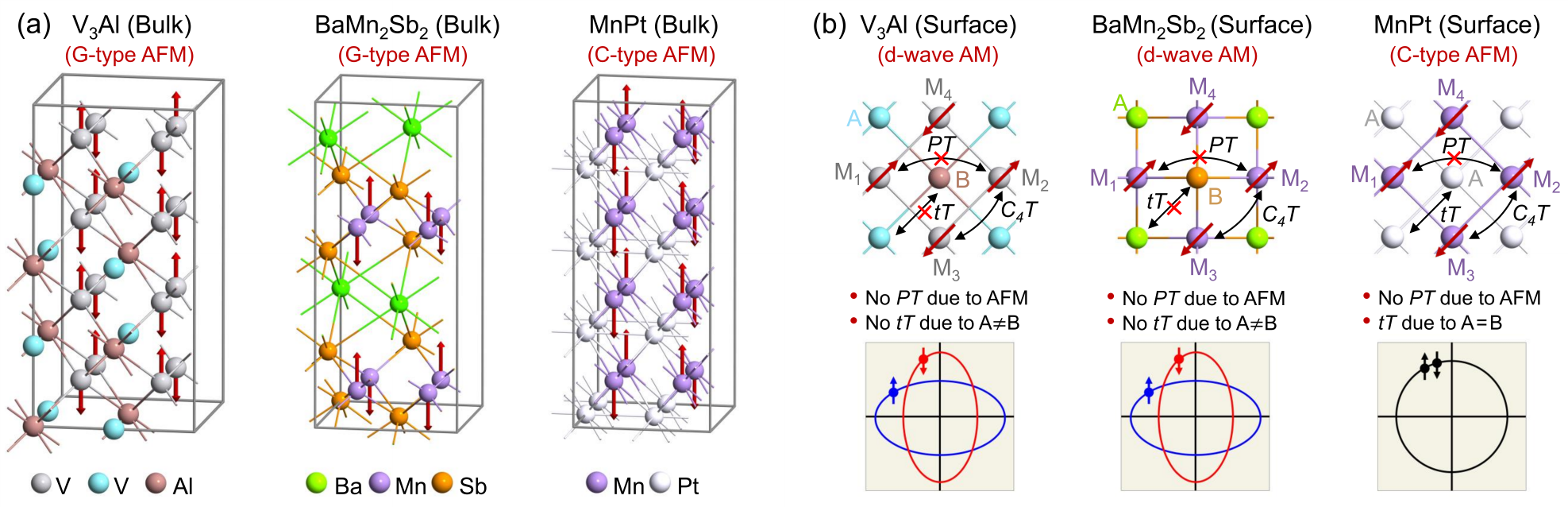}
\vspace{-0.6 cm}
\caption{Bulk and surface magnetic structures and symmetry conditions.
(a) Bulk magnetic configurations of the centrosymmetric AFMs V$_3$Al and BaMn$_2$Sb$_2$
(\textit{G}-type) and MnPt (\textit{C}-type). All bulk phases preserve combined $PT$ symmetry,
enforcing spin-degenerate electronic bands in the absence of SOC. Arrows indicate magnetic
moment orientations.
(b) Top view of the surface and subsurface layers at a representative surface termination.
The surface is taken along the (001) direction of the bulk unit cell, for which each atomic layer
contains two antiferromagnetically coupled magnetic atoms. For V$_3$Al and BaMn$_2$Sb$_2$, the
subsurface layer exhibits a checkerboard arrangement of inequivalent nonmagnetic sites,
making the local environments of the two antiferromagnetically coupled surface magnetic
sublattices inequivalent. As a result, no antiunitary symmetry remains that exchanges the two
surface sublattices. Although a fourfold rotation combined with time reversal ($C_4T$) remains,
this symmetry does not exchange the two surface sublattices and therefore does not enforce
spin degeneracy, allowing a symmetry-permitted nonrelativistic \textit{d}-wave altermagnetic
spin splitting, schematically indicated by the surface Fermi surfaces. In contrast, the MnPt
surface preserves a sublattice-exchanging antiunitary symmetry, realized as $tT$ symmetry,
despite broken inversion and therefore remains spin-degenerate. The indicated symmetry
operations summarize the distinct symmetry constraints in each case.}
\label{fig1}
\end{figure*}

In this Letter, we show that surface inversion breaking in centrosymmetric antiferromagnets,
while necessary, is not sufficient to induce surface altermagnetism. Instead, no antiunitary
symmetry may survive at the surface that exchanges the two antiferromagnetically coupled sublattices
and enforces spin degeneracy. Using symmetry analysis and first-principles calculations, we
demonstrate this mechanism for representative centrosymmetric AFMs including the Heusler compound
V$_3$Al and the 122 pnictide BaMn$_2$Sb$_2$. In these systems, the absence of a sublattice-exchanging
antiunitary symmetry allows a symmetry-permitted \textit{d}-wave momentum-dependent spin splitting
localized within the outermost magnetic layers.  We further show that the mechanism applies across multiple classes of centrosymmetric antiferromagnets and remains robust against SOC, although relativistic spin mixing can reduce the observable spin polarization in heavier-element compounds.

For surface terminations satisfying these symmetry conditions, the leading nonrelativistic spin splitting transforms according to a \textit{d}-wave representation determined by the surface magnetic point group \cite{Smejkal2022_PRX,Mazin2023_MnTe}. First-principles calculations confirm that this symmetry-allowed spin splitting emerges at realistic antiferromagnetic surfaces, exhibits characteristic momentum dependence, and remains strongly localized within the outermost magnetic layers. These results establish centrosymmetric AFMs as a previously unexplored platform for
nonrelativistic surface and interface altermagnetism. In contrast to bulk altermagnets, where
spin splitting is constrained by crystal rotations or mirror symmetries combined with time
reversal,  surface altermagnetism is governed by whether a sublattice-exchanging antiunitary symmetry 
survives at the surface. In systems such as MnPt this role is played by $tT$ symmetry.

To elucidate the symmetry mechanism responsible for this effect, Fig.~\ref{fig1} summarizes the
origin of surface altermagnetism by contrasting bulk and surface magnetic structures of
representative centrosymmetric AFMs. All materials considered are collinear \textit{C}- or
\textit{G}-type AFMs that preserve combined $PT$ symmetry in the bulk.
As shown in Fig.~\ref{fig1}(a), the bulk magnetic structures of V$_3$Al, BaMn$_2$Sb$_2$, and MnPt are
fully compensated and antiferromagnetically ordered. In the following, we focus on the (001)
surface, for which each magnetic layer contains two antiferromagnetically coupled magnetic moments.

At a surface, inversion symmetry is inherently broken and the combined $PT$ symmetry is therefore
absent even for centrosymmetric AFMs. However, as illustrated in Fig.~\ref{fig1}(b), the absence of
$PT$ alone is not sufficient to generate altermagnetic spin polarization. Spin degeneracy can still
be enforced by additional symmetries that relate opposite-spin magnetic moments within the surface
plane. In particular, spin degeneracy survives if the surface magnetic structure preserves an
antiunitary symmetry that exchanges the two antiferromagnetic sublattices. In systems such as
MnPt, this symmetry is realized as  $tT$ symmetry, where a half
translation $t$ exchanges the two sublattices.

The role of these symmetries is made explicit in the top views of the surface and subsurface layers
shown in Fig.~\ref{fig1}(b). The magnetic atoms are labeled M$_1$ and M$_2$ (spin-up) and M$_3$ and
M$_4$ (spin-down), while nonmagnetic atoms in the surface unit cell are denoted by A and B. For
V$_3$Al and BaMn$_2$Sb$_2$, the surface termination breaks inversion symmetry, as indicated by the
absence of a $PT$ operation that maps the two spin-up magnetic moments M$_1$ and M$_2$ onto each
other. Moreover, the subsurface layer exhibits a checkerboard arrangement of chemically distinct
nonmagnetic sites A and B, resulting in different local environments for the surface magnetic atoms.
Importantly, surface altermagnetism is governed not by generic sublattice inequivalence alone, but 
by the absence of a sublattice-exchanging antiunitary symmetry.
Although a fourfold rotation combined with time reversal ($C_{4}T$) remains and maps M$_3$ onto
M$_2$, this symmetry does not enforce spin degeneracy because it does not exchange the two
spin-polarized surface sublattices. Consequently, the surface electronic states are free to develop
a nonrelativistic momentum-dependent spin splitting characteristic of surface altermagnetism.

In contrast, for MnPt the nonmagnetic surface sites A and B are equivalent (both corresponding to Pt
atoms), such that the local environments of the surface magnetic atoms remain identical.
Consequently, a half translation exchanging the two sites remains a symmetry of the surface.
Accordingly, a sublattice-exchanging antiunitary symmetry survives, realized here as 
$tT$ symmetry, despite the absence of inversion symmetry.
The survival of this symmetry enforces spin degeneracy of the surface electronic structure,
making MnPt a conventional AFM at the surface with no altermagnetic spin splitting.
This example highlights that surface inversion breaking alone does not guarantee surface
altermagnetism; the presence or absence of a sublattice-exchanging antiunitary symmetry is decisive.

The symmetry condition identified above has direct implications for materials selection.
Table~\ref{tab1} illustrates its generality by surveying representative compounds from major
classes of centrosymmetric collinear AFMs considered in this work (an extended list is provided in
the Supplemental Material \cite{SM}). The representative systems considered here illustrate the 
general symmetry mechanism in experimentally relevant antiferromagnets, including Heusler compounds, 
122 pnictides, and related layered materials with compensated magnetic surface structures favorable 
for surface-confined altermagnetic states. All listed materials possess centrosymmetric bulk crystal 
structures and  preserve combined $PT$ symmetry in the bulk.  Surface altermagnetism emerges only 
when no sublattice-exchanging antiunitary symmetry survives at the surface. In systems such as MnPt, 
where this role is played by $tT$ symmetry, spin degeneracy remains protected. Importantly, the emergence of surface altermagnetism is determined not by magnetic order type alone, but by the specific antiunitary and rotational symmetries that survive at a given surface termination.
In materials such as MnPt, LaFeO$_3$, and K$_2$NiF$_4$, a half translation remains that exchanges equivalent surface sites, preserving the degeneracy-protecting
symmetry and enforcing spin degeneracy despite broken inversion symmetry. By contrast, LaMnPO lacks 
such a surviving sublattice-exchanging symmetry and therefore hosts surface altermagnetism despite 
having the same \textit{C}-type magnetic order as MnPt. This comparison illustrates explicitly that 
surface altermagnetism is governed not by magnetic order type alone, but by the specific symmetry that 
survives at the surface termination. While the presence or absence of a degeneracy-protecting symmetry 
determines whether the effect occurs, the magnitude of the spin splitting, when allowed, is material-dependent 
and reflects the electronic character of the surface states. Together, Fig.~\ref{fig1} and Table~\ref{tab1}
establish the absence of sublattice-exchanging antiunitary symmetries at the surface as the defining
condition for surface altermagnetism in collinear AFMs and provide a practical guideline for identifying 
candidate materials.

\begin{table}[t]
\caption{Representative centrosymmetric collinear AFMs illustrating the symmetry
condition for surface altermagnetism (see Supplemental Material \cite{SM} for an
extended survey of candidate materials across multiple structural families).
All materials preserve combined $PT$ symmetry in the bulk and therefore exhibit
spin-degenerate bulk electronic bands in the absence of SOC. Surface altermagnetism
(AM) is symmetry-allowed only when no antiunitary symmetry $\Sigma$ remains at the
surface that exchanges the two antiferromagnetically coupled sublattices. The column
``$\Sigma$ present'' indicates whether such a sublattice-exchanging symmetry survives
at the surface (e.g., $\Sigma=tT$ in MnPt).} 
\label{tab1}
\begin{ruledtabular}
\begin{tabular}{llccc}
Material & Class & AFM type & $\Sigma$ present & AM \\
\hline
V$_3$Al          & L2$_1$ Heusler        & G-type & No  & Yes \\
BaMn$_2$Sb$_2$   & 122 pnictide          & G-type & No  & Yes \\
MnPt             & L1$_0$ intermetallic  & C-type & Yes ($tT$) & No  \\
LaMnPO           & 1111 oxypnictide      & C-type & No & Yes \\
LaFeO$_3$        & Perovskite oxide      & G-type & Yes & No  \\
K$_2$NiF$_4$     & Layered perovskite    & G-type & Yes & No \\
\end{tabular}
\end{ruledtabular}
\end{table}

\begin{figure*}[t]
\centering
\includegraphics[width=0.9999\textwidth]{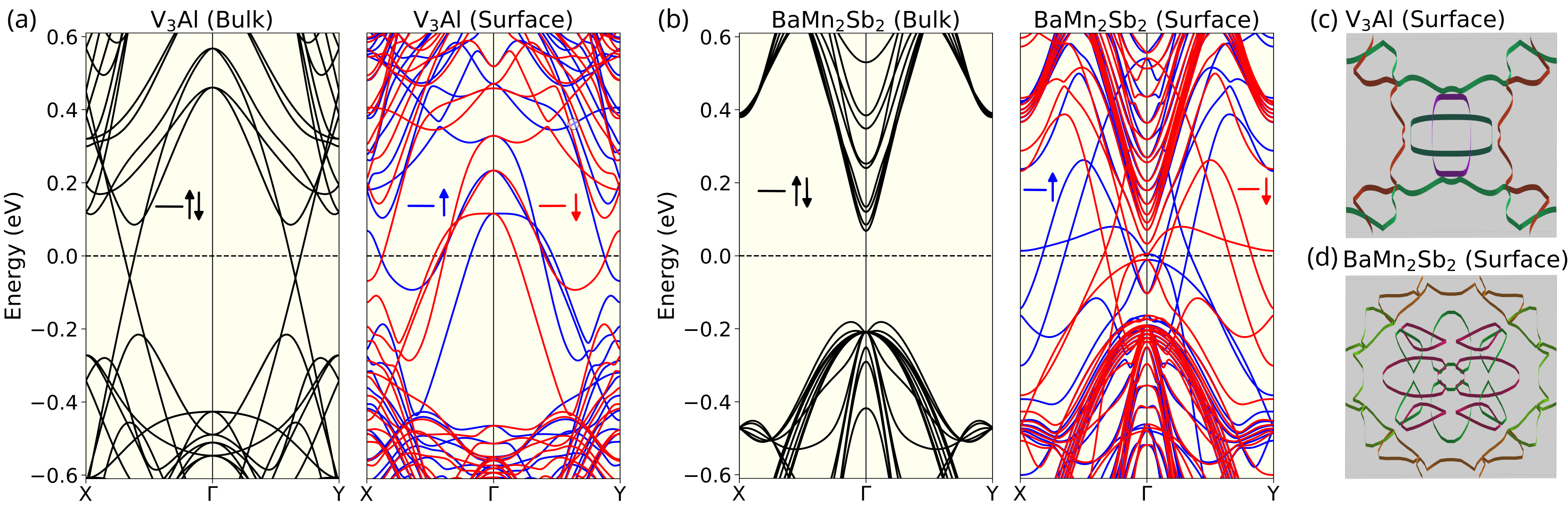}
\vspace{-0.6 cm}
\caption{Surface altermagnetic electronic structure of centrosymmetric AFMs.
(a) Bulk and surface band structures of the \textit{G}-type AFM V$_3$Al calculated along
the in-plane X--$\Gamma$--Y high-symmetry path. The surface band structure is obtained from a
20-layer slab, while the bulk reference is calculated using a corresponding 20-layer bulk unit cell
to enable direct comparison. The bulk bands are strictly spin-degenerate due to combined $PT$
symmetry, whereas the surface bands exhibit pronounced nonrelativistic spin
splitting in surface-localized states once no degeneracy-protecting symmetry remains
that exchanges the two antiferromagnetic surface sublattices.
(b) Same as (a), but for the \textit{G}-type AFM BaMn$_2$Sb$_2$.
(c), (d) Spin-resolved two-dimensional Fermi surfaces of the full slabs for V$_3$Al and
BaMn$_2$Sb$_2$, respectively. The characteristic four-lobe angular dependence reflects 
the \textit{d}-wave symmetry of the altermagnetic surface states.}
\label{fig2}
\end{figure*}

We now test this symmetry-based criterion at the microscopic level using first-principles electronic-structure 
calculations for representative AFMs that satisfy the surface symmetry conditions for altermagnetism,
namely the absence of a degeneracy-protecting symmetry that exchanges the two antiferromagnetic surface
sublattices. We focus on the G-type AFMs V$_3$Al and BaMn$_2$Sb$_2$, identified in Fig.~\ref{fig1} and Table~\ref{tab1}
as prototypical systems in which surface altermagnetism is symmetry-allowed. The calculations are based on density-functional 
theory within the generalized gradient approximation, treating magnetism in the collinear and nonrelativistic 
limit \cite{smidstrup2019an,perdew1996generalized,QuantumATKPseudoDojo}. Surface electronic structures are obtained 
from slab geometries constructed along the (001) surface orientation. In symmetric slabs, the two opposing surface 
terminations are related by the underlying G-type antiferromagnetic order, such that partner surface states on 
opposite sides exhibit opposite spin polarizations. As a result, their spin splittings compensate in the total 
spectrum, preserving global spin degeneracy. While the underlying splitting can always be recovered from surface- 
or layer-resolved projections, for clarity of presentation we therefore employ asymmetric slab geometries in which 
the altermagnetic splitting appears directly in the total band structure. Corresponding results for symmetric slabs, 
as well as further computational details, are provided in the Supplemental Material \cite{SM}. Unless stated otherwise, SOC is neglected in order to isolate the nonrelativistic spin splitting originating from the antiferromagnetic exchange field. Additional SOC calculations are presented below and in the Supplemental Material \cite{SM}.

Figure~\ref{fig2}(a) shows the bulk and surface band structures of V$_3$Al along the in-plane
X--$\Gamma$--Y high-symmetry path, using an orthorhombic surface Brillouin zone with $a=b$.
V$_3$Al is a high-$T_N$ antiferromagnet, with reported Néel temperatures of the order of
600~K, and has been discussed as a spin-gapless AFM
\cite{Galanakis2016_V3Al,khmelevskyi2016first,tas2017design,chen2019effects}.
In the bulk reference calculation, the electronic bands are strictly spin-degenerate and exhibit
a Dirac-like dispersion near the X point, with the Dirac node located approximately 60~meV below the
Fermi level.  At the surface, where no degeneracy-protecting symmetry remains, pronounced 
nonrelativistic spin splitting emerges in states localized
near the outermost magnetic layers. The Dirac cones shift toward the X and Y points and are pinned
close to the Fermi level, with each linear branch carrying opposite spin polarization.  The resulting 
altermagnetic spin splitting reaches values of up to 0.32~eV, giving rise to
metallic Dirac surface states with a strong momentum-dependent spin polarization.

An electronically distinct manifestation of surface altermagnetism is found in BaMn$_2$Sb$_2$,
shown in Fig.~\ref{fig2}(b). BaMn$_2$Sb$_2$ is a layered antiferromagnetic
semiconductor with a small bulk band gap of about 0.3~eV and a reported Néel temperature of
approximately 450~K \cite{An2009,Sangeetha2017,Zhang2019}. While the bulk electronic structure
remains fully spin-degenerate, the surface band structure develops two Dirac points located
approximately 20~meV above the Fermi level along the $\Gamma$--X and $\Gamma$--Y directions.
In the absence of SOC, these Dirac states are fully spin-polarized in a momentum-selective manner: 
bands along $\Gamma$--X carry one spin character, whereas those along $\Gamma$--Y carry the opposite spin.
The resulting altermagnetic spin splitting reaches values of up to $\sim$0.35~eV near the Fermi
level, demonstrating that pronounced nonrelativistic spin polarization can emerge even at the
surface of a small-gap antiferromagnetic semiconductor, in sharp contrast to its spin-degenerate
bulk electronic structure.

The anisotropic spin polarization of the surface bands discussed above has a direct manifestation
in momentum space at the Fermi level. These momentum-space signatures of surface altermagnetism are
illustrated in Fig.~\ref{fig2}(c) and Fig.~\ref{fig2}(d), which show spin-resolved two-dimensional
surface Fermi surfaces obtained from asymmetric slab calculations. In BaMn$_2$Sb$_2$, only two
surface-derived bands cross the Fermi level, giving rise to a comparatively clean multi-lobed
Fermi surface with alternating spin polarization. In contrast, several surface bands contribute
near the Fermi level in V$_3$Al, resulting in a more complex Fermi-surface texture; for clarity,
only two representative bands per spin channel are shown. Despite these material-specific
differences, both systems exhibit a pronounced fourfold angular modulation with alternating spin
character, providing a direct momentum-space signature of the \textit{d}-wave symmetry imposed by
the surface magnetic structure.

\begin{figure}[t]
\centering
\includegraphics[width=\columnwidth]{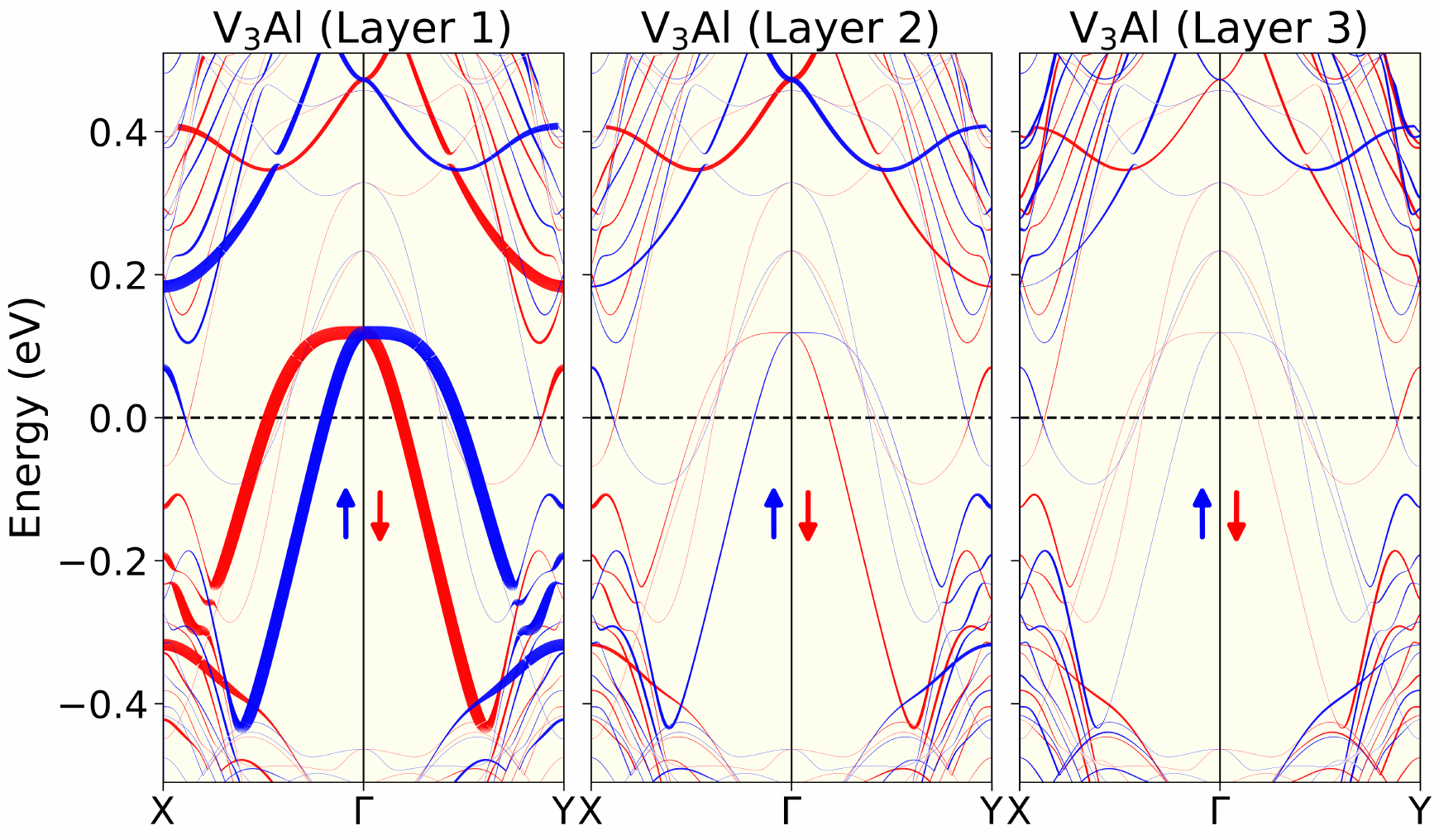}
\vspace{-0.5 cm}
\caption{
Layer-resolved surface altermagnetic states in V$_3$Al.
Fat-band representations of the asymmetric 20-layer V$_3$Al slab projected onto the first three
magnetic V layers, shown along the in-plane X--$\Gamma$--Y path.
The surface magnetic layer (Layer~1) hosts pronounced nonrelativistic \textit{d}-wave altermagnetic
spin splitting near the Fermi level with dominant spectral weight.
The same split surface-derived states persist in the subsurface layers (Layers~2 and~3) with rapidly
decreasing weight.}
\label{fig3}
\end{figure}

Having established the symmetry origin and momentum-space signatures of surface altermagnetism,
we now turn to its spatial localization. Figure~\ref{fig3} presents layer-resolved fat-band
representations for the asymmetric V$_3$Al slab, projected onto the first three magnetic V layers
of the V--V terminated top surface. The nonrelativistic \textit{d}-wave spin splitting near the
Fermi level is dominated by hole-like surface-derived states localized in the outermost magnetic
layer, which carries the majority of the spectral weight. The same split surface bands persist in
the subsurface layers with rapidly diminishing weight, while their dispersion and splitting remain
essentially unchanged. In deeper layers, surface-state contributions become negligible and the
electronic structure recovers the $PT$-protected spin degeneracy characteristic of the bulk.
Because the surface states are metallic and partially delocalized, a unique layer-resolved energy
splitting cannot be defined; instead, the progressive loss of spectral weight provides a direct and
physically transparent measure of the decay of surface altermagnetism into the slab interior.

To exclude finite-slab effects associated with the asymmetric geometry, we also performed
semi-infinite surface calculations for the V--V terminated V$_3$Al(001) surface. As shown in the
Supplemental Material~\cite{SM}, the semi-infinite surface spectral function reproduces the same
hole-like spin-split surface bands observed in the slab calculations, confirming that the
characteristic \textit{d}-wave splitting is intrinsic to the V--V surface termination rather than a
consequence of slab thickness or charge redistribution between opposite surfaces. Additional
layer-resolved projections for both the top and bottom terminations of V$_3$Al and
BaMn$_2$Sb$_2$ are also presented in the Supplemental Material. In both compounds, the
magnetic-atom-terminated top surfaces exhibit strongly localized spin-split states confined
essentially within the first surface unit cell. By contrast, the nonmagnetic-atom-terminated
bottom surfaces display more extended surface states with substantial \textit{d}-wave spin splitting
penetrating more deeply into the slab interior. The corresponding surface dispersions further
exhibit distinct electron- and hole-like character depending on the surface termination,
highlighting the strong termination dependence of the surface electronic structure.

To examine the robustness of surface altermagnetism against relativistic effects, we additionally
performed SOC calculations for both V$_3$Al and BaMn$_2$Sb$_2$ (see Supplemental Material
\cite{SM}). In V$_3$Al, SOC produces only negligible modifications of the nonrelativistic
\textit{d}-wave spin splitting due to the relatively weak atomic SOC of the constituent elements.
In contrast, for the MnSb-terminated surface of BaMn$_2$Sb$_2$, SOC induces noticeable spin
mixing and additional splitting of crossing surface bands, reducing the spin polarization while
preserving the characteristic \textit{d}-wave momentum dependence of the surface states. For the
Ba-terminated surface, however, SOC effects remain weak. Additional noncollinear SOC calculations
did not reveal any surface spin canting in either V$_3$Al or BaMn$_2$Sb$_2$, consistent
with the dominant nearest-neighbor antiferromagnetic exchange interactions compared to relativistic
SOC-induced interactions. These results demonstrate that surface altermagnetism remains robust in
the presence of SOC, while the observable spin polarization can depend sensitively on relativistic
spin mixing in heavier-element compounds.

The symmetry criterion identified here provides a practical route for extending surface
altermagnetism to a broad range of centrosymmetric AFMs. Additional candidate materials
spanning multiple structural families are listed in the Supplemental Material \cite{SM}. The
strong surface localization and nonrelativistic origin of the effect further suggest that analogous
states should emerge at antiferromagnetic interfaces and heterostructures where inversion
symmetry is locally broken. Surface altermagnetism should therefore be accessible to
surface-sensitive probes such as angle-resolved photoemission spectroscopy and scanning
tunneling microscopy. While the present work focuses on idealized surface terminations,
additional structural relaxations discussed in the Supplemental Material did not reveal any
tendency toward surface reconstruction in the considered systems. Surface roughness,
reconstruction, and disorder may nevertheless quantitatively modify the surface electronic
structure and spin splitting in real materials. Nevertheless, the local altermagnetic spin
polarization is expected to remain robust as long as the relevant magnetic arrangement and
surface symmetry conditions are approximately preserved. Together, these results establish
surface altermagnetism as a versatile platform for exploring symmetry-controlled spin
phenomena at antiferromagnetic surfaces and interfaces.

In summary, we establish surface altermagnetism as a symmetry-controlled phenomenon emerging 
at the surfaces of centrosymmetric collinear AFMs. Combining symmetry analysis with first-principles 
calculations, we show that surface inversion breaking alone is insufficient to induce momentum-dependent 
spin splitting. Instead, the effect arises only when no sublattice-exchanging antiunitary symmetry 
survives at the surface. We demonstrate this mechanism explicitly for V$_3$Al and BaMn$_2$Sb$_2$,
which host robust \textit{d}-wave altermagnetic surface states despite fully spin-degenerate bulk 
electronic structures, while MnPt remains spin degenerate due to the survival of $tT$ symmetry. The 
resulting states are strongly localized within the outermost magnetic layers and originate from 
nonrelativistic exchange-driven spin splitting that emerges already in the absence of SOC. 
Explicit SOC calculations further show that the characteristic \textit{d}-wave momentum dependence 
remains robust, although relativistic spin mixing can reduce the observable spin polarization 
in heavier-element compounds. These results establish a general framework for identifying and 
engineering altermagnetic surfaces and interfaces, substantially expanding the materials platform 
for momentum-dependent spin polarization in antiferromagnets.

\emph{Note added.}—
After submission of this work, we became aware of a complementary study based on surface spin-group analysis and database screening of candidate materials for surface altermagnetism \cite{lange2026emergent}.

\begin{acknowledgments}
This work was supported by Deutsche Forschungsgemeinschaft (DFG): project 328545488 – CRC/
TRR 227, Project No. B12 and LO 1659/10-1. 
\end{acknowledgments}

%\section*{Data Availability Statement}

%Data available on request from the authors

%\nocite{*}
\bibliography{surface_altermagnetism}

\end{document}